# Effect of the Sc/Zr ratio on the corrosion resistance of Al-Mg cast alloys

N.A. Kozlova, V.N. Chuvil'deev, A.V. Nokhrin[(*)], Y.S. Shadrina, A.A. Bobrov, M.K. Chegurov

Lobachevsky State University of Nizhniy Novgorod, 603022 Russia, Nizhny Novgorod, Gagarina ave., 23

e-mail: chuvildeev@nifti.unn.ru

**Abstract**

The results of investigations of the corrosion resistance of Al-Mg-Sc-Zr alloys with varying Mg content and different Sc/Zr ratios are presented. The objects of investigations were the Al-Mg-Sc-Zr alloys with total Sc + Zr content of 0.32 wt.%. The concentration of Sc and Zr in the alloys varied with the increments of 0.02 wt.%. The alloys were produced by induction casting. The effect of annealing temperature on the microhardness and electrical resistivity of the Al-Mg-Sc-Zr alloys was investigated. Corrosion tests were carried out in a medium simulating intergranular corrosion in aluminum alloys. Electrochemical studies and mass loss tests were performed. An increase in the Sc concentration and a decrease in the Zr one were shown to result in an increase in the corrosion rate. The primary $Al_3(Sc_xZr_{1-x})$ particles were found to have the main effect on the corrosion resistance of Al-Mg-Sc-Zr alloys. The dependence of the corrosion current on the annealing temperature of the Al-Mg-Sc-Zr alloy was found to have a non-monotonous character (with a maximum).

**Highlights**

- The resistance of Al-(2.5, 4, 6)%Mg alloys with different Sc/Zr ratios to intergranular corrosion (IGC) was investigated
- Increasing the Sc concentration and decreasing the Zr one lead to an increase in the IGC rate of Al-Mg alloys.

---

[(*)] Corresponding author (nokhrin@nifti.unn.ru)

- The formation of primary Al$_3$(Sc,Zr) particles accelerates the corrosion of Al-Mg alloys.
- A non-monotonous character of the dependence of corrosion current on the annealing temperature of Al-Mg-Sc-Zr alloys was observed.

1. **Introduction**

Al-Mg alloys possess an optimal combination of strength, ductility, and corrosion resistance that causes their wide use in mechanical engineering, shipbuilding, and aviation technology. Traditionally, the problem of increasing the strength and corrosion resistance of aluminum is solved by choosing an optimal composition of alloying elements (AEs), among which scandium is one of the most efficient AEs for aluminum alloys [1, 2]. The high cost of Sc limits the prospects of using Sc-containing aluminum alloys significantly, making it relevant to replace Sc with cheaper rare earth elements (REEs) or transition metals.

Zirconium is one of the most efficient AEs allowing partial replacement of Sc in Al-Mg alloys. The optimal ratio of Sc/Zr to achieve high strength in Al-Mg alloys was found to be ~3:1 (wt.%). This has led to the development of a group of industrial Al-Mg-Sc-Zr alloys (grades 1570, 1570C, 1571, 1575, etc.) with increased strength [3]. The investigations on optimizing the Sc/Zr ratio continue, and it is worth mentioning studies aimed at creating the alloys with Sc/Zr = 1 (wt.%) [4-9], which are directed towards further reducing the concentration of expensive Sc in the Al alloys.

Currently, there is no consensus among the researchers how the Sc/Zr ratio affects the corrosion resistance of aluminum alloys. The analysis of the literature shows that Sc and Zr can have an ambiguous impact on the corrosion resistance of aluminum alloys, and the data presented in the literature are contradictory.

In [10-12], it is shown that there may be a relationship between the corrosion rate of aluminum alloy and the grain size similar to the Hall-Petch relation. The precipitation of Al$_3$X particles allows forming a fine-grained microstructure, which in turn can increase the corrosion resistance of aluminum (see, for example, [13-16]). In [17], it was demonstrated that in Al-Zn-Mg alloys with additions of Sc and Zr, the precipitation of Al$_3$(Sc,Zr) particles contributes to an increase in the volume fraction of high-angle grain boundaries (HAGBs). In fine-grained Al-Mg and Al-Mg-Mn alloys with additions of Sc and Zr, a decrease in the volume fraction of the β-phase Mg$_2$Al$_3$ particles located at the grain boundaries may be observed [18-20]. The formation of a fine-grained structure allows reducing the tendency of aluminum alloys to form grain boundary segregations of corrosion-prone elements (Zn, Mg, Cu), thereby enhancing their resistance to local corrosion attack (see [17, 21-24]). In [25], it was noted that in the alloys with additions of Sc and Zr, smaller Al$_6$(Mn,Fe) particles were observed, which affect the corrosion resistance of aluminum positively. The positive effect of Sc on the corrosion resistance and stress corrosion cracking resistance of aluminum alloys has been reported in previous studies [20, 26-29, etc.]. Thus, the Sc/Zr ratio affects the tendency for recrystallization and grain growth in Al-Mg alloys (see, for example, [30, 31]) and allows forming a fine-grained microstructure with a large extent of grain boundaries. This helps reducing the local concentration of detrimental elements at grain boundaries, allowing reducing the size and changing the distribution pattern of the particles along grain boundaries. These factors affect the corrosion resistance of aluminum alloys positively, primarily their resistance to various forms of localized corrosion.

Some authors express the opposite opinion regarding the nature of the effect of Sc and Zr on the corrosion resistance of aluminum alloys. It is known that the accelerated formation and growth of secondary particles can occur in the HAGBs, especially when the grain boundaries of the alloy contain an increased concentration of AEs. In this case, the formation of the fine-grained structure with a high fraction of HAGBs in Al-(Sc,Zr) alloys may lead to a reduction in the corrosion resistance of the alloy (see, for example, [22, 23, 32]).

Sometimes, it is noted that HAGBs have an increased energy and, as a result, low corrosion resistance even in high-purity aluminum [21]. The falling of defects (dislocations, vacancies) into the grain boundaries can lead to an additional reduction in the corrosion resistance of aluminum alloys [33, 34]. It is worth noting the article [35], which highlights that the addition of Sc and Zr allows increasing the resistance of Al-Mg alloys to intergranular corrosion (IGC), but the authors point out the need to consider the type of grain boundaries – maximum resistance to IGC was observed in the case of subgrain structure formation. Similar result was demonstrated in [18].

The presence of $Al_3(Sc,Zr)$ particles may have a negative effect on the corrosion resistance of aluminum alloys due to the formation of microgalvanic couples between the Al crystal lattice - $Al_3(Sc,Zr)$ particle (see [36-38]). The studies such as [39-41] have shown that $Al_3Sc$ and $Al_3(Sc, Zr)$ particles act as cathodes with respect to the aluminum crystal lattice. As a result, the increase in corrosion resistance associated with the refinement of the aluminum grain structure may be offset by the negative electrochemical behavior of $Al_3(Sc,Zr)$ particles. In [32], it is noted that the addition of 0.25% Sc did not lead to a significant change in the corrosion resistance of aluminum alloy 7010 but was accompanied by the appearance of an additional (second) breakdown potential on the anodic polarization curve and a shift of the corrosion potential into the active region. This increased the susceptibility of 7010 alloy to IGC and pitting corrosion. Similar ambiguous results were obtained in other studies [29, 42, 43].

In [44, 45], the dependence of the corrosion rate of Al-(2.5-3)%Mg-X%Sc-(0.14-0.15)%Zr alloy on the scandium concentration (X = 0, 0.15, 0.3, 0.6, and 0.9%Sc) was demonstrated to have a non-monotonous character with a minimum corresponding to 0.3% Sc. It is suggested that the $Al3(Sc_xZr_{1-x})$ particles may dissolve in neutral aqueous environments (3% NaCl) that leads to the formation of a passivating layer containing scandium and $Sc_2O_3$ oxide on the aluminum surface. It is known also that the resistance of the aluminum oxide film can be increased by increasing the Sc concentration [46] including through annealing of Al-4%Mg-0.3%Sc and Al-4%Mg-0.24%Sc-0.06%Yb alloys [47]. Presumably, the negative electrochemical effect of $Al_3(Sc,Zr)$ particles on the corrosion resistance of aluminum begins to prevail over the positive effect of scandium on the resistance of the protective aluminum oxide film with a substantial increase in the Sc and Zr concentrations.

The addition of Sc and Zr allows refining the as-cast structure of the aluminum alloy by forming a large number of coarse primary $Al_3(Sc,Zr)$ particles during solidification (see [48]). As mentioned above, this can lead to a reduction in the fraction of grain boundaries occupied by the β-phase $Mg_2Al_3$ particles. This enables an increase in the corrosion resistance of the alloy without the need for prolonged additional homogenizing annealing, which is crucial for achieving high strength in Al-Mg-Sc-Zr alloys. The temperature and time of homogenizing annealing for Al-Mg alloys [49] noticeably differ from the optimal temperature-time regime of dispersion strengthening for Al-Mg-(Sc,Zr) alloys due to the precipitation of $Al_3(Sc,Zr)$ nanoparticles. High-temperature homogenizing annealing will result in rapid coarsening of the $Al_3(Sc,Zr)$ particles and a decrease in the strength of Al-Mg-(Sc,Zr) alloys. On the other hand, the application of prolonged low-temperature homogenizing annealing often proves to be insufficient in providing high corrosion resistance for Al-Mg-(Sc,Zr) alloys (see, for example, [50]).

The objective of the present study was to investigate the effect of the Sc/Zr ratio on the corrosion resistance of Al-Mg alloys. The fundamental hypothesis underlying this study was that changing the Sc/Zr ratio in the solution zed Al-Mg alloy alters the average grain size of the alloy and, consequently, the number and nature of the distribution of β-phase particles located along the grain boundaries of aluminum. This primarily affects the susceptibility of aluminum alloys to IGC. Furthermore, the change in the Sc/Zr ratio may also affect the composition of primary $Al_3(Sc_xZr_{1-x})$ particles and, therefore, the susceptibility of solution zed alloys to pitting corrosion. Annealing of Al-Mg-Sc-Zr alloys leads to the precipitation of $Al_3(Sc,Zr)$ particles within the crystalline lattice; the composition of the precipitating particles affects the tendency of the aluminum alloy to the localized (pitting) corrosion.

## 2. Materials and Methods

The work focuses on Al-Mg-Sc-Zr alloys with total Sc + Zr content of 0.32 wt.%. The concentrations of Sc and Zr in the alloys varied with an increment of 0.02 wt.%. The magnesium concentration in the alloys was 2.5, 4.0, and 6.0 wt.%. The composition of the Al alloys is presented in Table 1.

Table 1. Compositions of aluminum alloys

| Series | Alloy # | Alloying elements, % wt. (%at.) | | | | |
|---|---|---|---|---|---|---|
| | | Mg | Sc | Zr | Sc+Zr | Sc/Zr |
| 1 | 1-2.5 | 2.5 (2.8) | 0.10 (0.060) | 0.22 (0.065) | 0.32 (0.125) | 0.45 (0.92) |
| | 1-4.0 | 4.0 (4.4) | | | | |
| | 1-6.0 | 6.0 (6.7) | | | | |
| 2 | 2-2.5 | 2.5 (2.8) | 0.12 (0.072) | 0.20 (0.059) | 0.32 (0.131) | 0.60 (1.22) |
| | 2-4.0 | 4.0 (4.4) | | | | |
| | 2-6.0 | 6.0 (6.7) | | | | |
| 3 | 3-2.5 | 2.5 (2.8) | 0.14 (0.084) | 0.18 (0.053) | 0.32 (0.137) | 0.78 (1.58) |
| | 3-4.0 | 4.0 (4.4) | | | | |
| | 3-6.0 | 6.0 (6.7) | | | | |
| 4 | 4-2.5 | 2.5 (2.8) | 0.16 (0.096) | 0.16 (0.047) | 0.32 (0.143) | 1.00 (2.04) |
| | 4-4.0 | 4.0 (4.4) | | | | |
| | 4-6.0 | 6.0 (6.7) | | | | |
| 5 | 5-2.5 | 2.5 (2.8) | 0.18 (0.108) | 0.14 (0.041) | 0.32 (0.149) | 1.29 (2.63) |
| | 5-4.0 | 4.0 (4.4) | | | | |
| | 5-6.0 | 6.0 (6.7) | | | | |
| 6 | 6-2.5 | 2.5 (2.8) | 0.20 (0.120) | 0.12 (0.035) | 0.32 (0.156) | 1.67 (3.43) |
| | 6-4.0 | 4.0 (4.4) | | | | |
| | 6-6.0 | 6.0 (6.7) | | | | |
| 7 | 7-2.5 | 2.5 (2.8) | 0.22 (0.132) | 0.10 (0.030) | 0.32 (0.162) | 2.20 (4.40) |
| | 7-4.0 | 4.0 (4.4) | | | | |
| | 7-6.0 | 6.0 (6.7) | | | | |

The workpieces of aluminum alloys with the dimensions of 20×20×160 mm were obtained by vacuum induction casting using the Indutherm VTC 200V casting machine with a ceramic crucible. Argon purging was performed before melting the components and during heating. The molten metal was also subjected to electromagnetic stirring to enhance its chemical homogeneity. The melting time of the components ranged from 6 to 10 min at 800 °C. The casting temperature was maintained at 760 °C. The molten metal was held at this temperature for 60 min prior to casting. Cooling down was performed within 50–250 s with vibration. The workpieces were solidified in a copper mold. The alloys were made from A99-grade aluminum, Mg90-grade magnesium, and master alloys Al-10%Zr and Al-2%Sc, which were produced by vacuum induction casting followed by rolling into 0.2 mm foil. No homogenizing annealing of the cast blanks was performed.

The microstructure of the alloys was examined using an optical microscope Leica DM IRM and a scanning electron microscope (SEM) Jeol JSM-6490 equipped with an Oxford Instruments INCA 350 EDS microanalyzer. The microhardness and electrical resistivity measurements were employed to study the precipitation of $Al_3(Sc,Zr)$ particles. Microhardness (Hv) was measured using a Qness A50+ hardness tester with a 50 g load. The average uncertainty in Hv determination was ±15 MPa. The specific electrical resistivity (SER) of the alloys was measured using the eddy current method with a SIGMATEST 2.069 instrument and a sensor of 8 mm in diameter.

Corrosion studies were carried out in at room temperature, in a 3% NaCl + 0.3% HCl aqueous solution (pH = 1.18). The tests were performed using R-20X and R-30S potentiostats in standard three-electrode glass cell of 60 mL in volume. A silver-chloride reference electrode (EVL-1M4) and a platinum electrode served as the auxiliary electrode were used in testing. Samples with the dimensions of 2×10×15 mm were polished using diamond pastas to achieve a surface roughness level of 3-5 μm, then coated with acid-proof lacquer, except open area of ~1 $cm^2$. Before starting the electrochemical tests, the samples were pre-exposed in the test cell in the corrosion solution for 1 h until a stable open circuit potential was reached while simultaneously monitoring the potential-time dependence E($t$). During the 1-hour period, the potential of the aluminum alloy reached its stationary value. The investigation was carried out within the potential range from -1.0 to 0.4 V with a potential scanning rate of 0.5 mV/s and a recording rate of 10 points/s. The ES8 software from Elin Co was used for result analysis. The corrosion current was determined from Tafel curves. Based on the analysis of the slope in the Tafel curves [the potential $E$ vs. the current $i$ in the semilogarithmic axes lg($i$)–$E$], the corrosion current $i_{corr}$ and the corrosion potential $E_{corr}$ were calculated. To check the repeatability of the results, not less than three samples were tested for each structural-phase state. Mean uncertainty of determining $E_{corr}$ was 5 mV and was limited by the reproducibility of the results. The uncertainty of determining the $i_{corr}$ varied since it depended on the uncertainty of determining the open corrosion area, on the uncertainty of determining the slope in the Tafel curves by least mean squares method, and on the reproducibility of the results. Mean uncertainty of determining $i_{corr}$ was close to 0.1–0.15 mA/cm$^2$.

Gravimetric tests were carried out at room temperature. Corrosion studies were carried out in an aqueous solution of 3% NaCl + 0.3% HCl (pH = 1.18) was performed according to the requirements of Russian National Standard GOST 9.021-74. The investigations have shown the Al-Mg alloys are to be destroyed in this solution via the IGC mechanism. The testing time was 24 h. Prior to the testing, the surfaces of the specimens underwent mechanical polishing down to the roughness level of 3–5 μm. Samples 2×10×15 mm were placed in a test corrosion cell in a suspended state. The geometric dimensions of the samples were measured with an accuracy of 0.1 mm. The mass of the samples was determined using Sartorius CPA 225D analytical scales. The accuracy in measuring the mass of the samples was ±0.0005 g. The classification of the emerging corrosion defect type was performed according to the requirements of Russian National Standard GOST 9.908-85.

### 3. Results

#### 3.1 Microstructure of Alloys

The cast alloys exhibited a homogeneous fine-grained macrostructure with a thin layer of columnar crystals at the edges of the cross-sections of the ingots (Fig. 1). An increase in Mg content leads to nearly complete disappearance of the columnar crystals and a reduction in the average grain size in the central parts of the ingots. In alloys with 6% Mg, the average grain sizes in the central parts of the ingots were close to 30-50 μm (Fig. 2).

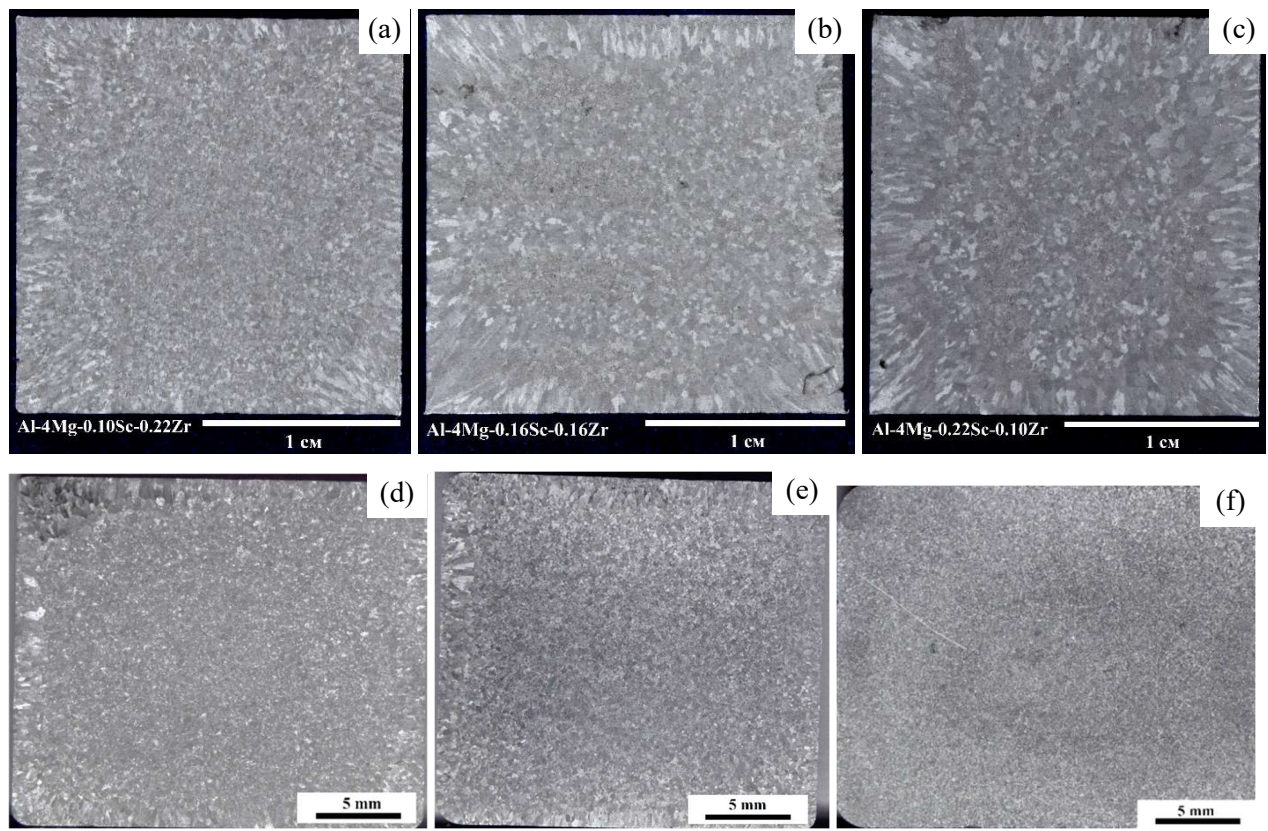

**Fig. 1**. Macrostructure of samples with 4% Mg (a, b, c) and 6% Mg (d, e, f) with different Sc/Zr ratios: (a, d) Sc/Zr = 0.45 (wt.%), (b, e) Sc/Zr = 1.0 (wt.%); (c, f) Sc/Zr = 2.2 (wt.%).

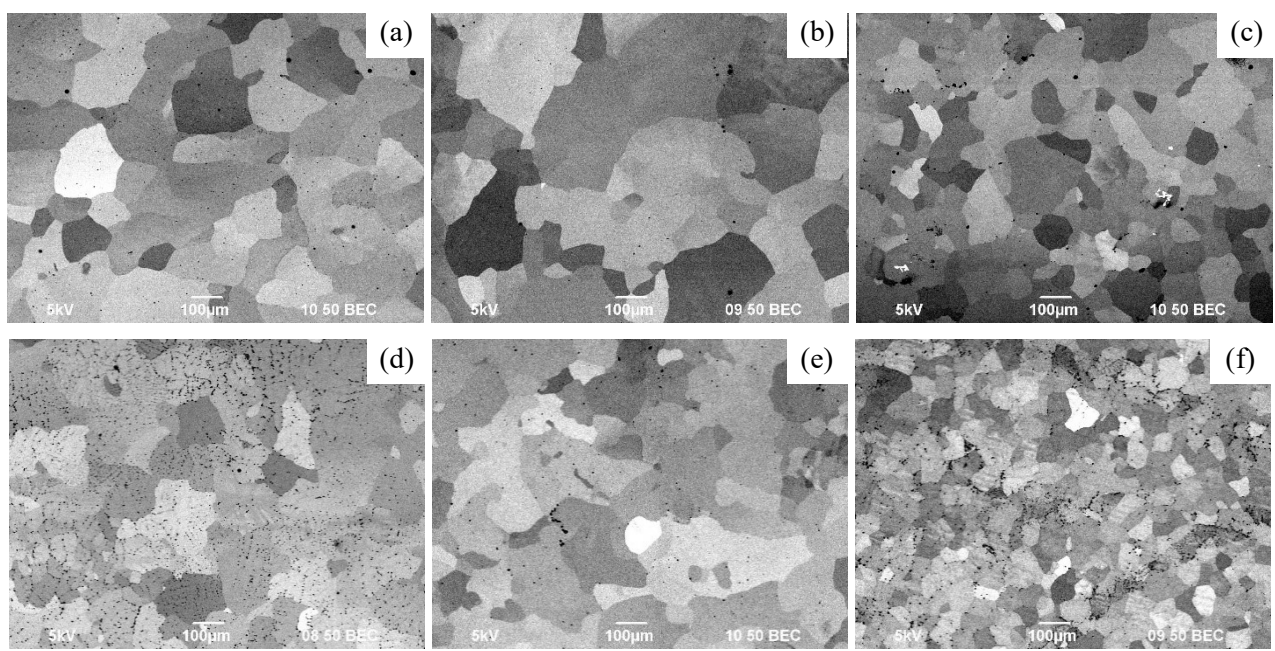

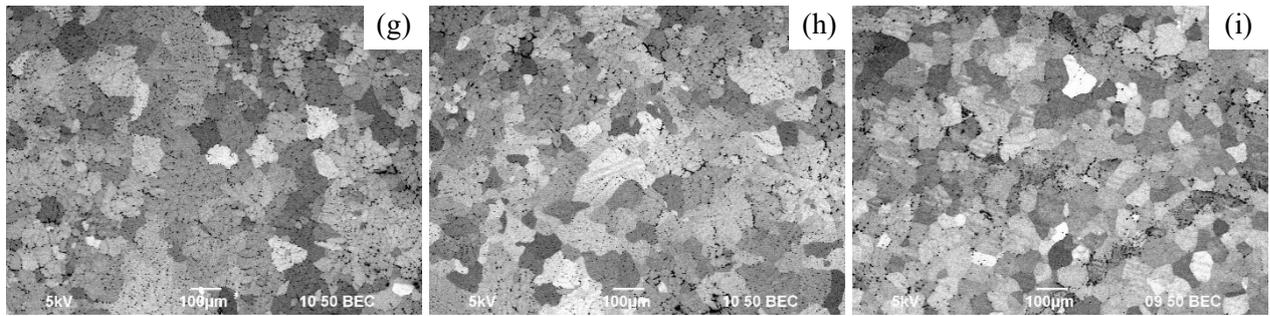

**Fig. 2**. Microstructure of the central parts of the samples with 2.5% Mg (a, b, c), 4% Mg (d, e, f), and 6% Mg (g, h, i) with different Sc/Zr ratios: (a, d, g) Sc/Zr = 0.45 (wt.%), (b, e, h) Sc/Zr = 1.0 (wt.%); (c, f, i) Sc/Zr = 2.2 (wt.%)

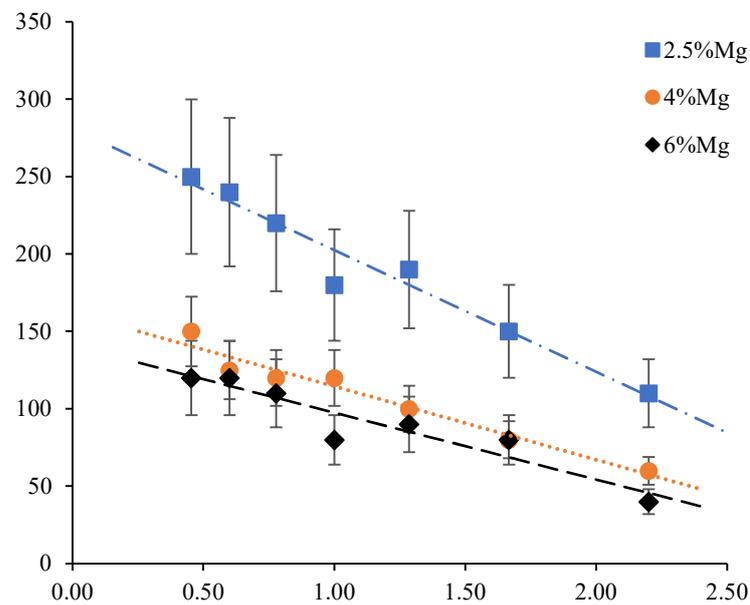

**Fig. 3**. Dependence of the average grain size in the central parts of the ingots on the Sc/Zr ratio

In the cast alloys, primary $Al_3(Sc,Zr)$ particles formed during crystallization were distributed uniformly on the surfaces of the cross-sectionals. In the Al-Mg-Sc-Zr alloys with elevated zirconium content (0.20, 0.22% Zr), large rectangular particles were observed inside small equiaxial grains (Fig. 4a, b). The EDS microanalysis has shown the composition of such particles to be close to $Al_3Zr$. In the cast alloys with increased scandium content, the isolated primary $Al_3(Sc,Zr)$ particles were distributed uniformly throughout the volume of the alloy (Fig. 4c, d).

An increase in the magnesium content leads to an increase in the volume fraction of the primary $Al_3(Sc,Zr)$ particles. The boundaries of the fine grains in the central parts of the ingots contained an increased amount of β-phase particles, which were destroyed intensively during electrochemical polishing (Fig. 4d, e). On the surfaces of strongly etched samples, an increased content of β-phase particles or an increased Mg content at the grain boundaries led to a change in the contrast of grain boundaries (Fig. 4b, d). An increase in the magnesium concentration in the Al-Mg alloys resulted in an increase in the volume fraction of β-phase particles.

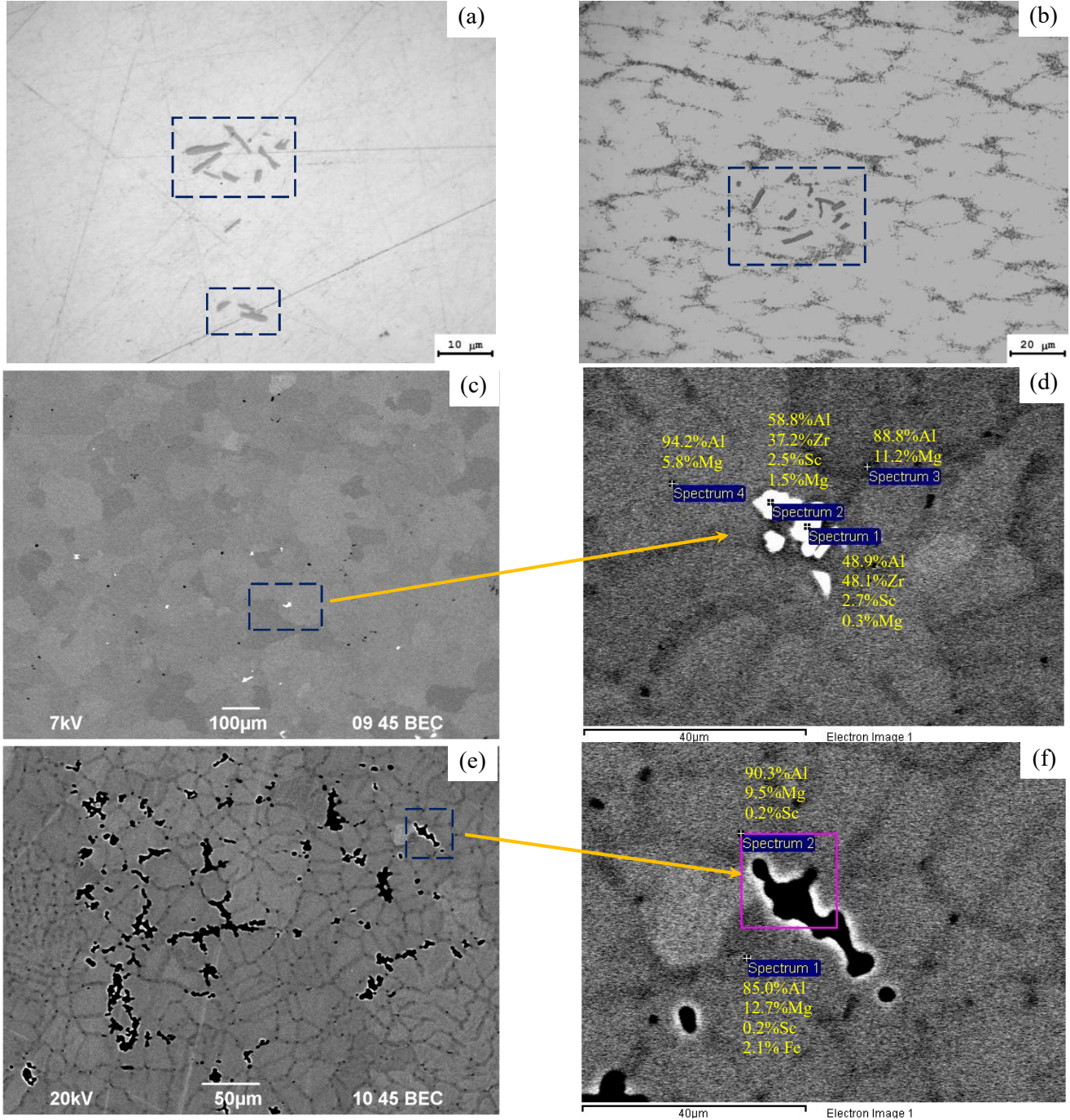

**Fig. 4**. Microstructure of Al-6%Mg-Sc-Zr alloys: (a, b) primary Al$_3$Zr particles in Al-6%Mg-0.10%Sc-0.22%Zr alloy; (c, d) primary Al$_3$(Sc,Zr) particles in Al-6%Mg-0.16%Sc-0.16%Zr alloy; (e, f) microstructure of Al-6%Mg-0.22%Sc-0.10%Zr alloy

### 3.2 Characterization of cast alloys

Table 2 presents the results of microhardness and SER studies of Al-Mg alloys in the initial state. An increase in the Mg concentration leads to an increase in Hv and SER of aluminum alloys at all Sc/Zr ratios. From Table 2, one can see that increasing the Sc concentration and decreasing the Zr one lead to a slight increase in SER in the alloys with 4 and 6% Mg; in the alloys with 2.5% Mg, changing the Sc/Zr ratio did not affect the SER significantly. It should be noted that for all alloys, the values of SER ($\rho_0$) measured experimentally was less than its theoretical value $\rho_{th}$, calculated assuming the additivity of the SER contributions to the SER of pure aluminum (see [51]). This result indicates that during the crystallization process, some AEs were segregated from the solid solution or metal melt leading to the formation of primary particles. The increased difference $\Delta\rho = \rho_{th}-\rho_0$ in Al-6%Mg-Sc-Zr alloys is associated with the formation of β-phase particles during the crystallization process and a decrease in the concentration of Mg in the aluminum crystal lattice.

**Table 2.** Results of investigations of properties of Al-Mg-Sc-Zr alloys in the initial state

|  | Al-2.5%Mg-Sc-Zr alloys | | | | | | |
|---|---|---|---|---|---|---|---|
| Alloy # | 1-2.5 | 2-2.5 | 3-2.5 | 4-2.5 | 5-2.5 | 6-2.5 | 7-2.5 |
| Hv, MPa | 580 ± 55 | 605 ± 40 | 595 ± 60 | 645 ± 60 | 625 ± 50 | 630 ± 30 | 635 ± 40 |
| $\Delta Hv_{max}$, MPa | 210 | 220 | 285 | 290 | 360 | 320 | 370 |
| $\rho_0$, µΩ·cm | 4.50 ± 0.04 | 4.59 ± 0.03 | 4.37 ± 0.03 | 4.62 ± 0.03 | 4.55 ± 0.02 | 4.54 ± 0.03 | 4.50 ± 0.03 |
| $\Delta\rho_{max}$, µΩ·cm | 0.30 | 0.33 | 0.35 | 0.38 | 0.34 | 0.39 | 0.38 |
| $\rho_{th}$, µΩ·cm | 4.59 | 4.60 | 4.60 | 4.61 | 4.61 | 4.62 | 4.62 |
|  | Al-4%Mg-Sc-Zr alloys | | | | | | |
| Alloy # | 1-4.0 | 2-4.0 | 3-4.0 | 4-4.0 | 5-4.0 | 6-4.0 | 7-4.0 |
| Hv, MPa | 630 ± 40 | 660 ± 45 | 685 ± 40 | 690 ± 50 | 700 ± 30 | 655 ± 40 | 640 ± 40 |
| $\Delta Hv_{max}$, MPa | 235 | 255 | 285 | 300 | 345 | 365 | 390 |
| $\rho_0$, µΩ·cm | 5.25 ± 0.04 | 5.37 ± 0.04 | 5.34 ± 0.05 | 5.36 ± 0.05 | 5.37 ± 0.05 | 5.34 ± 0.04 | 5.39 ± 0.06 |
| $\Delta\rho_{max}$, µΩ·cm | 0.25 | 0.35 | 0.33 | 0.32 | 0.44 | 0.35 | 0.40 |

| $\rho_{th}$, μΩ·cm | 5.30 | 5.31 | 5.31 | 5.32 | 5.32 | 5.33 | 5.33 |
|---|---|---|---|---|---|---|---|
| | Al-6%Mg-Sc-Zr alloy | | | | | | |
| Alloy # | 1-6.0 | 2-6.0 | 3-6.0 | 4-6.0 | 5-6.0 | 6-6.0 | 7-6.0 |
| Hv, MPa | 730 ± 50 | 800 ± 30 | 795 ± 50 | 790 ± 50 | 780 ± 60 | 775 ± 60 | 770 ± 60 |
| $\Delta Hv_{max}$, MPa | 320 | 300 | 360 | 385 | 405 | 410 | 440 |
| $\rho_0$, μΩ·cm | 6.27 ± 0.04 | 6.30 ± 0.04 | 6.30 ± 0.05 | 6.32 ± 0.05 | 6.33 ± 0.06 | 6.34 ± 0.04 | 6.35 ± 0.05 |
| $\Delta\rho_{max}$, μΩ·cm | 0.10 | 0.16 | 0.16 | 0.17 | 0.17 | 0.17 | 0.22 |
| $\rho_{th}$, μΩ·cm | 6.49 | 6.49 | 6.50 | 6.50 | 6.51 | 6.51 | 6.52 |

In Fig. 5, the dependencies of the change in SER and Hv on the temperature of the 30-min annealing of the cast Al-Mg-Sc-Zr alloys with different Mg contents are presented. From Fig. 5a, one can see that in the Al-2.5%Mg-Sc-Zr alloys, the decrease in the SER begins at 300 °C and was independent on the Sc/Zr ratio practically. The start of the precipitation of the $Al_3(Sc,Zr)$ particles leads to an increase in the hardness of the aluminum alloys. The maximum values of hardness were achieved after annealing at 350 °C. Further increase in the annealing temperature leads to a decrease in the hardness, which, according to the Orowan equation, is due to rapid growth of the secondary $Al_3(Sc,Zr)$ particles.

At the temperatures above 450°C, a slight increase in the SER was observed, possibly, due to partial dissolution of the β-phase particles and an increase in the concentration of Mg in the aluminum crystal lattice. Support for this assumption comes from the consistent increase in the SER at the temperatures above 450 °C, which did not depend on the concentrations of Sc and Zr in the alloy within the uncertainty of the SER measurements. Similar trends were observed in the ρ(T) and Hv(T) dependencies for the alloys with 4%Mg (Fig. 5b) and 6%Mg (Fig. 5c).

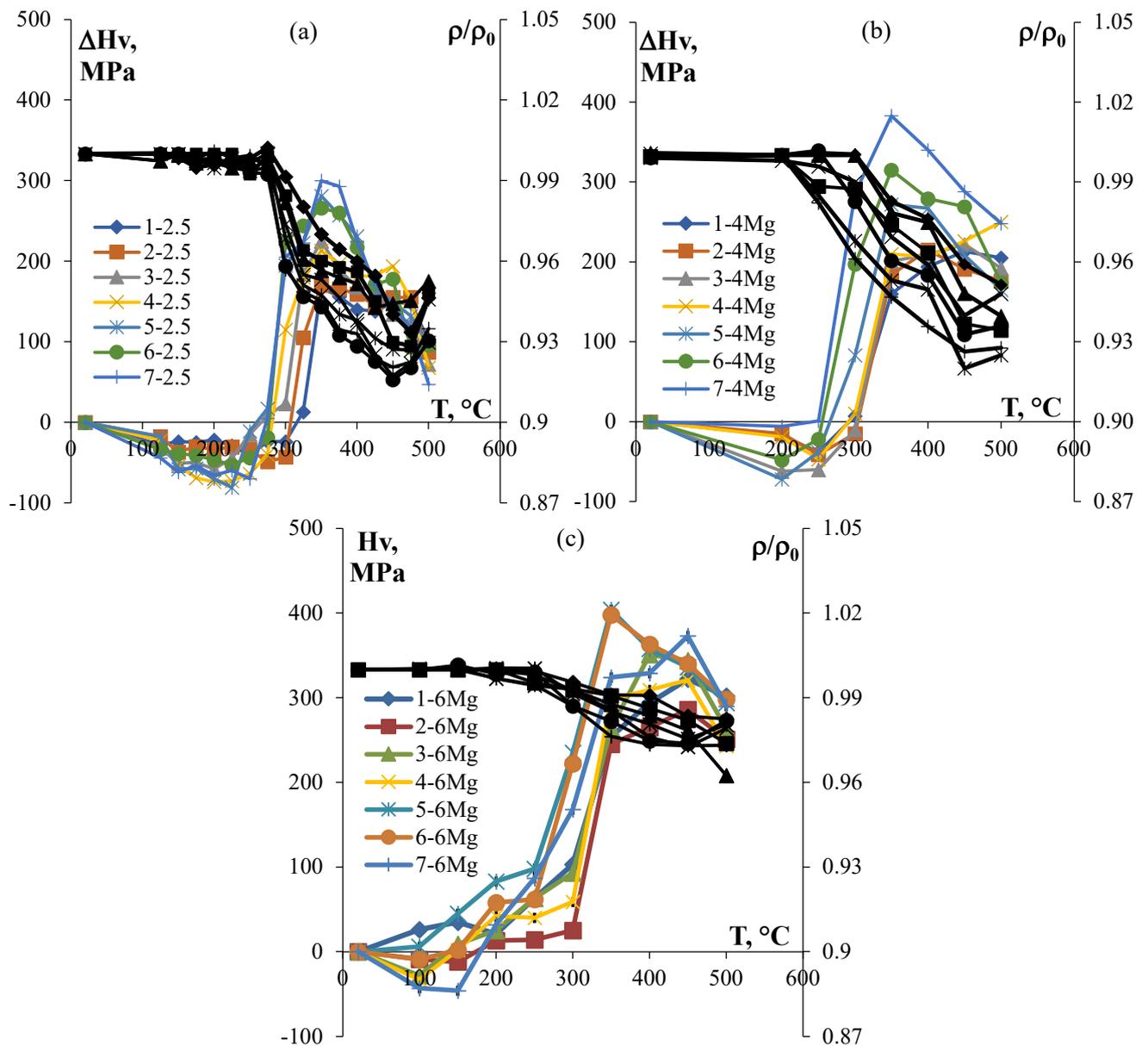

**Fig. 5**. Dependencies of the change in the SER ($\rho/\rho_0$, black lines) and microhardness ($\Delta Hv$, colored lines) on the temperature of annealing of the alloys with 2.5%Mg (a), 4%Mg (b), and 6%Mg (c)

The intensity of secondary $Al_3(Sc,Zr)$ particles' precipitation upon heating the Al-Mg-Sc-Zr alloys can be characterized by the increase in hardness ($\Delta Hv_{max}$) and the decrease in electrical resistivity ($\Delta\rho_{max} = \rho_0 - \rho_{min}$) during the heating process. From Table 2, one can see an increase in $\Delta Hv_{max}$ and $\Delta\rho_{max}$ with increasing Sc content (increasing Sc/Zr ratio). The result obtained indicates the alloys with higher Sc content to exhibit more intense precipitation of $Al_3(Sc,Zr)$ particles. An increase in the Mg concentration leads to a decrease in $\Delta\rho_{max}$ and a slight increase in $\Delta Hv_{max}$. In the Al-6%Mg-0.10%Zr-0.22%Sc alloy (alloy #7-6.0), the hardness during annealing increment reaches

440 MPa whereas in alloys #7-4.0 and #7-2.5, these were 390 and 370 MPa, respectively (Table 2). The decrease in $\Delta\rho_{max}$ indicates the increase in the Mg concentration to result in a reduction in the volume fraction of the secondary $Al_3(Sc,Zr)$ particles.

### 3.3 Corrosion Tests

#### 3.3.1 *Effect of Mg concentration and Sc/Zr ratio*

Fig. 6 shows the Tafel curves $lg(i) – E$ for the Al-Mg-Sc-Zr alloys with different Sc/Zr ratios. The Tafel curves $lg(i) – E$ had a conventional appearance and allow reliable determining the corrosion potential $E_{corr}$ and the corrosion current $i_{corr}$. For the ease of comparison, the $lg(i) – E$ curves for the alloys with 2.5, 4.0, and 6.0%Mg are presented in each plot. The analysis of the presented curves shows that increasing the magnesium concentration from 2.5 to 6% at a constant Sc/Zr ratio leads to a decrease in the $E_{corr}$ and an increase in the $i_{corr}$. The effect of the Mg concentration was most noticeable for the alloys with elevated Sc content.

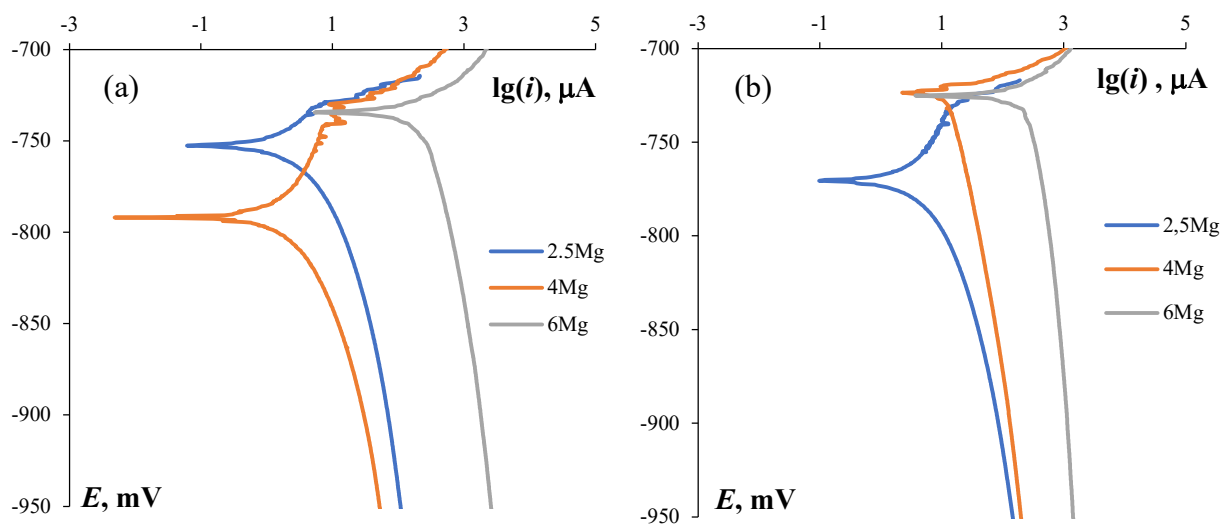

**Fig. 6**. Tafel curves for Al-Mg-Sc-Zr alloys with Sc/Zr ratio of 0.45 (wt.%) (a) and 2.2 (wt.%) (b). The blue line represents the alloy with 2.5%Mg, the red one - with 4%Mg, and the black one - with 6%Mg

Fig. 7 presents the dependence of the corrosion current $i_{corr}$ for the alloys in the initial unannealed state on the Sc/Zr ratio. One can see from Fig. 7 that increasing the concentration of Sc and decreasing the one of Zr lead to significant increase in the corrosion current $i_{corr}$ in the alloys with 4% and 6%Mg. In the alloys with 2.5%Mg, the change in the Sc/Zr ratio did not affect the corrosion current $i_{corr}$ noticeably.

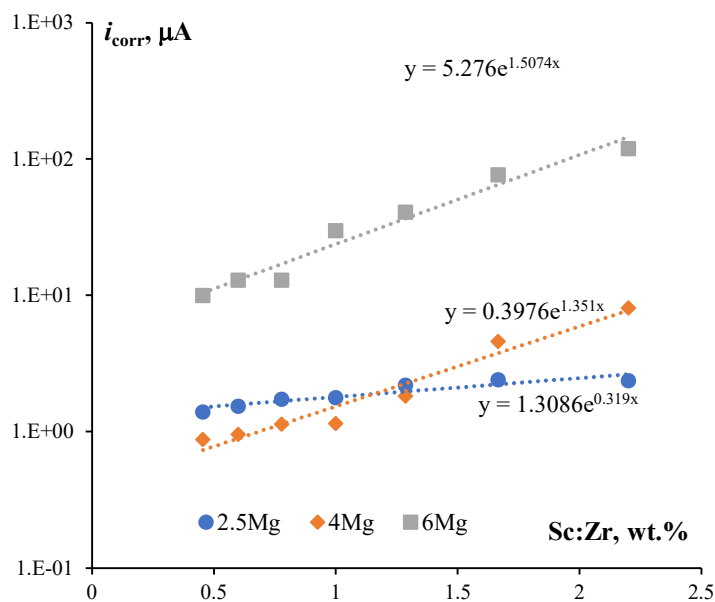

**Fig. 7**. Dependence of the corrosion current density on the Sc/Zr ratio in Al-Mg-Sc-Zr alloys

The results of the electrochemical studies correspond well to the ones of mass loss corrosion tests. In the alloys with 0.22%Sc + 0.10%Zr, an increase in magnesium concentration from 2.5 to 6% leads to an increase in the corrosion rate from 0.12 to 0.20 g·m$^{-2}$·h$^{-1}$. In the alloys with 0.16%Sc + 0.16%Zr and 0.10%Sc + 0.22%Zr, an increase in Mg concentration from 2.5 to 6% resulted in an increase in the corrosion rate from 0.13 to 0.28 g·m$^{-2}$·h$^{-1}$ and from 0.29 to 0.43 g·m$^{-2}$·h$^{-1}$, respectively. From Fig. 8, it is evident that the increase in magnesium content in the alloy alters the corrosion nature. In the alloys with 2.5%Mg, the isolated pitting corrosion was observed in the areas where β-phase particles were located (Fig. 8a). In the alloy with 6%Mg, the IGC was observed (Fig. 8b). The intergranular nature of corrosion in the alloys with 6%Mg is likely due to an increase in the number of β-phase particles along the grain boundaries in aluminum and an increase in the Mg concentration at the grain boundaries (Fig. 3).

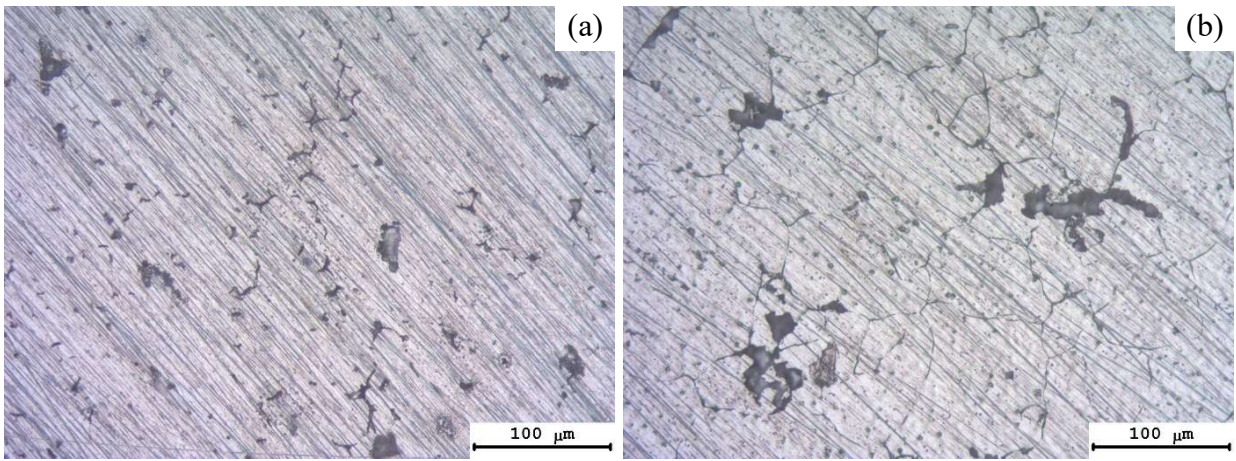

**Fig. 8**. Character of the corrosion damage of the surfaces of Al-2.5%Mg-0.22%Sc-0.10%Zr (a) and Al-6%Mg-0.22%Sc-0.10%Zr (b) alloy specimens after 24 h of mass loss corrosion testing. Metallographic optical microscopy

*3.3.2 Effect of annealing*

Fig. 9 presents typical Tafel curves for the specimens of Al-Mg-Sc-Zr alloys after annealing at various temperatures.

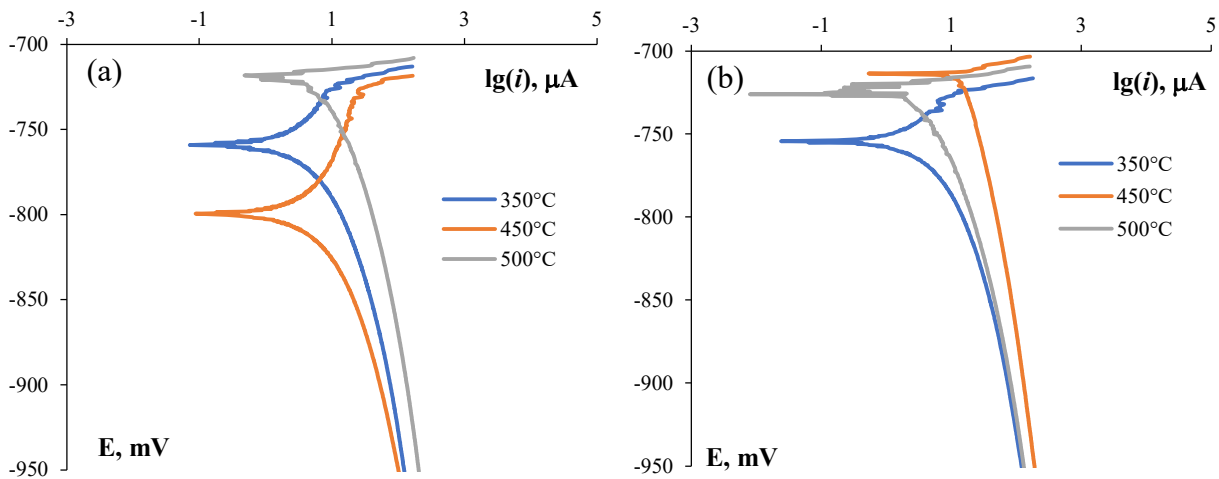

**Fig. 9**. Tafel curves for Al-2.5Mg-Sc-Zr alloys after annealing at 350, 450, and 500 °C: (a) Sc/Zr = 0.45 (wt.%); (b) Sc/Zr = 2.2 (wt.%)

A summary of the results of investigations is presented in Fig. 10. One can see from Fig. 10 that in the Al-2.5%Mg-Sc-Zr alloys, the increase in the corrosion current begins at 300 °C and was practically independent on the Sc/Zr ratio. The maximum values of the corrosion current $i_{corr}$ were achieved after annealing at 450 °C. It is important to note that the magnitude of the maximum corrosion current depends on the Sc/Zr ratio and increases with increasing Sc content. Further increase in the annealing temperature leads to a decrease in the corrosion current $i_{corr}$. A similar trends for the dependencies of the corrosion current on the annealing temperature was observed for the alloys with 4% and 6%Mg.

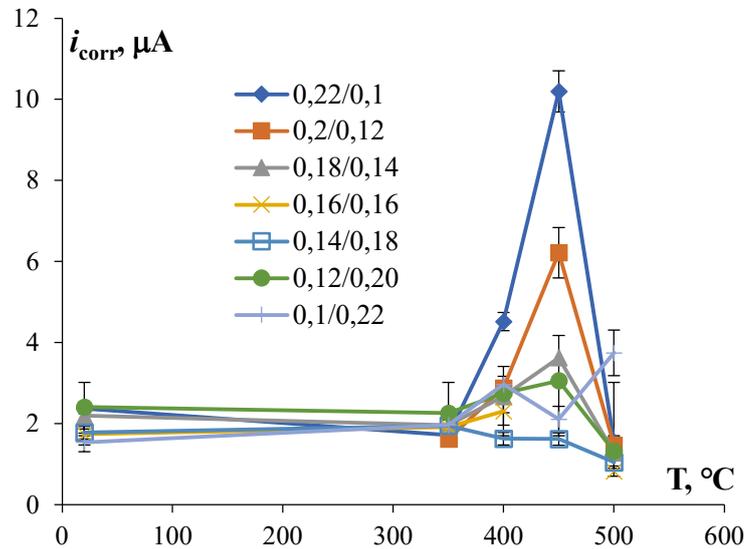

**Fig. 10.** Dependencies of the changes in the corrosion current on the annealing temperature for Al-2.5%Mg alloys with different Sc/Zr ratios

Annealing at 450°C did not affect the corrosion rate significantly – a slight decrease in the corrosion rate (mass loss) was observed, but the scale of the change in the corrosion rate exceeded the 'sample to sample' scatter of the properties of the investigated aluminum alloys only slightly. The nature of the corrosion damages of the surfaces of the annealed specimens remains unchanged. Main contribution to the failure of the alloys is made by the β-phase particles located along the grain boundaries (Fig. 11).

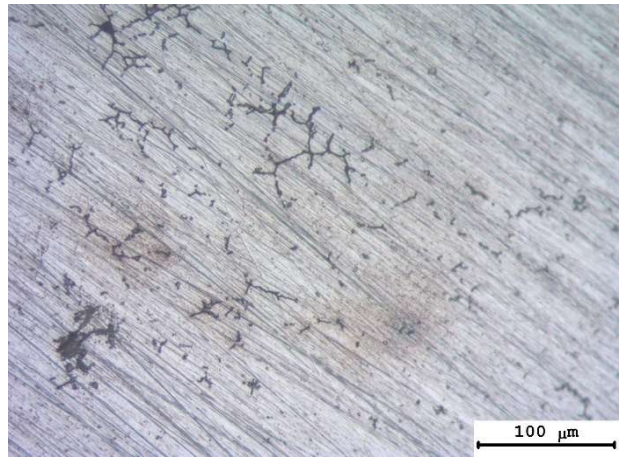

**Fig. 11.** Character of the corrosion damage of the surface of the Al-4%Mg-0.22%Sc-0.10%Zr specimen after annealing at 450°C for 30 min

## 4. Discussion

Let us analyze the effect of Mg concentration and of the Sc/Zr ratio on the corrosion rate of the Al-Mg-Sc-Zr alloys. It should be noted that the environment, which the corrosion tests of the aluminum alloys were carried out in was intended to investigate the IGC resistance.

As shown in Fig. 4 and 5, an increase in the magnesium concentration leads to an increase in the corrosion current. In the alloys with 6%Mg, the corrosion current exceeds that in the alloys with 2.5% and 4%Mg by an order of magnitude. In our opinion, the decrease in the corrosion resistance of the alloys with 6%Mg is due to the increase in the number of β-phase particles located along the grain boundaries. As shown above (Fig. 2), the β-phase particles exhibit low corrosion resistance and readily dissolve during electrochemical polishing of the ground surface. Additionally, the β-phase particles form microgalvanic couples with the aluminum crystal lattice, along the interphase boundaries of which an accelerated corrosion damage may occur.

The issue of the effect of the Sc/Zr ratio on the corrosion resistance of the Al-Mg alloys is less straightforward and requires further investigations to find out more reliable dependencies.

Thus, the result shows allow better understanding the effect of the Al-alloy composition on their corrosion resistance and highlights the importance of considering Mg content and the Sc/Zr ratio when designing the alloys with desired properties. As shown in Introduction, the $Al_3Sc$ particles act as the cathodes with respect to the aluminum crystal lattice and form the microgalvanic couples Al$_3$Sc/Al contributing to the acceleration of the localized corrosion. Therefore, one can assume the high values of $i_{corr}$ in the unannealed Al-Mg-Sc-Zr alloys with higher Sc/Zr ratios to originate from the presence of the primary Al$_3$(Sc,Zr) particles with an increased scandium content. The primary Al$_3$Zr particles and the Al$_3$(Sc,Zr) ones with higher zirconium content affect the corrosion current $i_{corr}$ less. Hence, partial substitution of scandium with zirconium (reducing the Sc/Zr ratio) leads to a decrease in the corrosion current (Fig. 5).

As one can see in Fig. 3, annealing results in a reduction of the SER of the alloys and an increase in their hardness. The changes observed were attributed to the reduction in the concentration of alloying elements (Sc, Zr) in the aluminum crystal lattice and in the formation of Al$_3$X particles, which hinder the dislocation movement. It is presumed that in the alloys with a higher scandium content (Sc/Zr > 1), the Al$_3$Sc particles and the variable-composition intermetallic compound Al$_3$(Sc$_x$Zr$_{1-x}$) ones with increased scandium content are segregated. According to [52, 53], these particles are more likely the Al$_3$(Sc$_{0.75}$Zr$_{0.25}$) ones. Since the Al$_3$Sc particles act as the cathodes with respect to the aluminum crystal lattice, the process of precipitation will lead to an increase in the corrosion current density. This assumption is supported by the fact that in the alloys with low scandium content and high zirconium one (Sc/Zr < 1), annealing has virtually no noticeable effect on the corrosion current magnitude (Fig. 7). This is likely due to the predominant formation of particles with a Al$_3$Sc core – Al$_3$Zr shell structure [54] during annealing as well as the Al$_3$(Sc$_{0.25}$Zr$_{0.75}$) particles (with increased zirconium content) [53, 54].

The decrease in the corrosion current of the Al-Mg-Sc-Zr alloys at higher annealing temperatures (above 450 °C), in our opinion, is attributed to the following two main factors.

First, at elevated heating temperatures, the dissolution of β-phase particles begins, and their contribution to the corrosion rate decreases. The indirect evidence of the beginning of the β-phase particle dissolution is the increase in the SER at elevated annealing temperatures (Fig. 7).

Second, the Al$_3$(Sc,Zr) particles begin to grow and larger particles absorb smaller ones during annealing. This leads to a reduction in the number of particles as well as a decrease in the Al / Al$_3$(Sc,Zr) interface area accompanied by a reduction in their contribution to the corrosion rate of the aluminum alloy.

The combination of these two factors leads to a reduction in the corrosion current for the Al-Mg-Sc-Zr alloys at elevated annealing temperatures.

The significant reduction in the corrosion current in the alloys with Sc/Zr > 1 (Fig. 7), in our opinion, is due to the specific structure of the Al$_3$(Sc,Zr) particles. As it was shown in [53, 54], the particles with a Al$_3$Sc core – Al$_3$Zr shell structure can be formed during the heating of Al-Sc-Zr alloys. Since the diffusion coefficient of Sc in Al is much larger than the one of zirconium [56-58], the formation of corrosion-prone Al$_3$Sc particles occurs initially during heating. Next, the Al$_3$Zr shells emerge on the surfaces of the Al$_3$Sc cores. This results in a reduction in the alloy's corrosion rate down to the values characteristic of the alloys with Sc/Zr > 1 with increased Zr content (Fig. 7).

## 5. Conclusions

1. The samples of Al-Mg-Sc-Zr aluminum alloys with total Sc + Zr content of 0.32 wt.% were obtained by induction casting. The concentrations of Sc and Zr in the alloys varied with a step of 0.02 wt.%. The Mg concentrations in the alloys were 2.5, 4.0, and 6.0 wt.%. The increasing of the Mg content was shown to result in an increase in the volume fraction of primary β-phase particles at the grain boundaries. In the alloys with increased Sc content, more intensive precipitation of the Al$_3$(Sc,Zr) particles was observed while an increase in the Mg concentration leads to a decrease in the fraction of the Al$_3$(Sc,Zr) particles. The effect of annealing temperature on the microhardness and SER of the cast Al-Mg-Sc-Zr alloys has been studied. Electrochemical corrosion tests have been carried out in an environment simulating IGC in the aluminum alloys.

2. The dependencies of changes in the SER and microhardness on the temperature of 30-minute annealing of the alloys, of the corrosion current for the alloys on the Sc/Zr ratio as well as on the temperature of 30-minute annealing have been analyzed. A decrease in the corrosion resistance of the alloys with 6% Mg in the corrosion environment designed to study the resistance of the aluminum alloys to the IGC was observed. In our opinion, this is associated with an increase in the number of the β-phase particles located along the grain boundaries. The assumption was made that high values of corrosion rate in the unannealed Al-Mg-Sc-Zr alloys with high Sc/Zr ratio are due to the presence of primary $Al_3(Sc,Zr)$ particles with increased scandium content in the alloys. Partial substitution of scandium with zirconium (reduction of the Sc/Zr ratio) leads to a decrease in the corrosion current. It was shown also that the process of precipitation leads to an increase in the corrosion current at the annealing temperatures up to 450 °C. At higher annealing temperatures, a decrease in the corrosion rate of the alloys was observed, that, in our opinion, is related to the dissolution of the β-phase particles as well as a decrease in the number of $Al_3(Sc,Zr)$ particles due to their growth and coalescence.

**Acknowledgments**

The authors would like to thank V.V. Zakharov (Russian Institute of Light Alloys, JSC, Moscow) for the recommendations on the selection of compositions and casting regimes of the aluminum alloys.